\documentclass[11pt,reqno]{amsart}

\usepackage[utf8]{inputenc}
\usepackage[letterpaper,margin=1in]{geometry}
\usepackage{amsmath,amssymb,amsthm,mathtools}
\usepackage{graphicx}
\usepackage{placeins}
\usepackage{booktabs}
\usepackage[round,authoryear]{natbib}
\usepackage{microtype}
\usepackage{titlesec}
\usepackage{setspace}
\usepackage{etoolbox}
\usepackage[colorlinks=true,linkcolor=blue,citecolor=blue,urlcolor=blue]{hyperref}
\usepackage{float}

\makeatletter
\patchcmd{\@settitle}
  {\uppercasenonmath\@title}
  {}
  {}
  {\PackageWarning{P1}{Could not patch title capitalization}}

\patchcmd{\@setauthors}
  {\MakeUppercase{\authors}}
  {\authors}
  {}
  {\PackageWarning{P1}{Could not patch author capitalization}}
\def\@captionfont{\normalfont\footnotesize}
\makeatother
\titleformat{\section}[hang]
  {\normalfont\large\bfseries}{\thesection}{0.8em}{}

\titlespacing{\section}
  {0pt}{3.5ex plus 1ex minus .2ex}{0.5ex}

\titleformat{\subsection}[hang]
  {\normalfont\bfseries}{\thesubsection}{0.8em}{}

\titlespacing{\subsection}
  {0pt}{2.5ex plus .8ex minus .2ex}{0.7ex}

\titleformat{\subsubsection}[hang]
  {\normalfont\itshape}{\thesubsubsection}{0.8em}{}

\titlespacing*{\subsubsection}
  {0pt}{2ex plus .5ex minus .2ex}{.8ex plus .2ex}
\numberwithin{equation}{section}

\theoremstyle{plain}
\newtheorem{theorem}{Theorem}[section]
\newtheorem{lemma}[theorem]{Lemma}

\newtheorem{corollary}[theorem]{Corollary}

\theoremstyle{definition}
\newtheorem{assumption}{Assumption}

\theoremstyle{remark}

\newtheorem*{remark*}{Remark}

\newcommand{\E}{\mathbb{E}}

\title[Renewable Estimation for GNR]{Renewable Estimation for Generalized Nonparametric Regression with Dynamic Spline Bases and Likelihood Pullback}

\author{Houtian Yi$^{1}$, Zhaolun Wang$^{1}$, Xin Zhao$^{1,*}$\\
\textnormal{\footnotesize
$^{1}$School of Mathematics, Southeast University, Nanjing, China}\\
\textnormal{\footnotesize
$^{*}$Corresponding author: xinzhaomath@seu.edu.cn}\\
\textnormal{\footnotesize
Contributing author: 230259082@seu.edu.cn, 220252001@seu.edu.cn}}

\date{} 

\begin{document}
\pagestyle{plain}
\raggedbottom
\begin{abstract}
A fixed spline space may limit approximation accuracy in generalized nonparametric regression with streaming data. We propose a renewable penalized-likelihood method for the nonparametric regression model based on canonical exponential-family. Our proposed method combines dynamic cubic-spline refinement with Gram-metric pullback of historical score and curvature summaries. The pullback preserves the retained historical likelihood quadratic on the embedded previous spline space, while incoming data update the fit and its summaries. Under regularity and algorithmic conditions, including a growing startup and an uncapped refinement regime, we establish $L^2$-consistency at rate $O_p(N^{-2/5})$ for a fixed target in $H^2[0,1]$ and a smaller-order energy-norm discrepancy from a matched offline estimator. Simulations and applications to the Voice and Catalina datasets show that dynamic bases improve estimation when the initial representation is restrictive, with additional gains from pullback. Comparison for the proposed method also shows lower computational and storage costs than the offline methods.

\end{abstract}


\maketitle

\section{Introduction}
\label{sec:intro}

Many applications in medicine and industry involve streaming datasets. We use the streaming setting to refer to data arriving in batches and the online setting to refer to observations arriving one at a time. Generally, both settings share similar ideas in statistical research. For example, both settings motivate sequential model updates without access to the full historical data, since repeatedly accessing the full dataset lead to much cost in computation and storage. Classical stochastic approximation \citep{robbinsmonro1951}, stochastic-gradient methods, and online convex optimization \citep{cesabianchi2021} provide approaches to update models as new observations arrive without access to the full dataset.

Related statistical research focus on updating parameter estimation and conducting inference as new batches arrive. For example, \citet{schifano2016} developed online updating algorithms for linear regression and estimating equations. \citet{luosong2020} introduced renewable estimation and incremental inference for generalized linear models. Their approach combines incoming likelihood contributions with summary statistics of previously processed observations, enabling sequential estimation and Wald test without accessing the full dataset. Particularly, the idea which is called renewable estimation inspires much research after such method based on GLM was proposed. 

Nonparametric methods can capture nonlinear relationships between a covariate and a response. In generalized nonparametric regression, the target is an unknown function, often estimated using penalized splines \citep{eilersmarx1996}. For streaming settings, With limited startup data, a relatively low-dimensional spline space is a practical choice for the initial model fitting. As data accumulate,  this space may be insufficient to represent the target function. Allowing the spline space to expand can alleviate this limitation. 

In streaming or online settings, nonparametric methods include smooth backfitting for generalized additive models \citep{yang2024}, sieve stochastic-gradient procedures and reproducing-kernel projection estimators \citep{zhangsimon2022,zhangsimon2023}, and one-pass penalized basis expansions under memory constraints \citep{quanlin2024}. Related work also includes renewable composite quantile regression \citep{chenfanglin2024} and online smoothing-parameter selection and adaptive estimation through rolling validation \citep{zhanglei2025,quanlong2026}. In particular, \citet{xueyao2022} developed a dynamic penalized-spline procedure in which sequential knot insertion generates nested spline spaces and historical summaries are updated through projection. Their work provides a foundation for dynamic spline bases in streaming settings. Offline convergence theory also provides results for penalized splines in generalized nonparametric regression and related likelihood-based models \citep{xiao2019,huangsu2021}.

In our algorithm, we aim to combine the idea of renewable estimation with dynamic bases for penalized-likelihood B-spline methods based on generalized nonparametric regression. Previously processed observations are compressed into a finite-dimensional historical summary: the fitted B-spline coefficient vector, cumulative score vector, and cumulative negative Hessian, also called curvature matrix. These quantities define a historical likelihood quadratic around the current fitted center. When the spline space expands, embedding the coefficients preserves the fitted function, but the score and curvature remain in the previous coordinates and require transport to the enlarged space.

To address this challenge, We use a Gram-metric pullback based on the Lebesgue-$L^2$ geometry of the receiving spline space. Composing the retained quadratic with the previous-space projection preserves it on the embedded previous space while defining its extension to the enlarged coordinates. The transported summary is then combined with the incoming likelihood and the current penalty. After each update, the score and curvature summaries are refreshed at the new fitted center. The algorithm supports updates in both cases—whether the spline space remains fixed (no basis expansion) or the basis is expanded, without access to the previous data.

The contributions are below in our study. First, we develop a renewable penalized-likelihood algorithm with dynamic spline-space refinement for the nonparametric regression model based on canonical exponential-family. Second, Gram-metric pullback gives a structured representation of historical quadratic information after expansion and preserves it on the embedded previous space. Third, under suitable assumptions, we establish $L^2$-consistency at rate $O_p(N^{-2/5})$ for a fixed target in $H^2[0,1]$ and a smaller-order energy-norm discrepancy from a matched offline fit. Simulations and applications to the Voice and Catalina datasets evaluate finite-sample estimation and prediction of our proposed method, with HTRU2 as a supplementary application. 

In the following part, Section~\ref{sec:method} introduces the model and renewable algorithm, and Section~\ref{sec:theory} presents the theoretical results. Sections~\ref{sec:simulation} and~\ref{sec:applications} report the simulation and real-data studies, followed by the discussion in Section~\ref{sec:discussion}.

\section{Problem Setup and Methodology}
\label{sec:method}
\subsection{GNR Model and Penalized Splines}
\label{subsec:gnr-splines}

Consider generalized nonparametric regression (GNR) with a scalar covariate mapped to $[0,1]$. Conditional on $X=x$, the response denoted as $Y$ follows a canonical one-parameter exponential family with density
\begin{equation}
\label{eq:edf}
f(y\mid x)=\exp\left\{\frac{yg_0(x)-b\{g_0(x)\}}{a_0}+c(y,a_0)\right\},
\qquad g_0(x)\in\Theta,
\end{equation}
where $a_0>0$ is known and fixed, $b$ is the cumulant function, and $\Theta\subseteq\mathbb R$ is the natural-parameter domain. The target is the natural-parameter function $g_0:[0,1]\to\Theta$, with
\begin{equation}
\label{eq:mean-variance}
\mu_0(x)=b'\{g_0(x)\},
\qquad \operatorname{Var}(Y\mid X=x)=a_0b''\{g_0(x)\}.
\end{equation}
Our simulations and applications to the real data use the Bernoulli logistic model, for which $b(\eta)=\log(1+e^\eta)$, $a_0=1$, and
\begin{equation}
\label{eq:logistic}
\mu_0(x)=\frac{\exp\{g_0(x)\}}{1+\exp\{g_0(x)\}}.
\end{equation}
Other examples include Gaussian responses with $b(\eta)=\eta^2/2$ and known variance $a_0$, Poisson responses with $b(\eta)=e^\eta$, and binomial responses with $r$ fixed trials and $b(\eta)=r\log(1+e^\eta)$; the latter two both have $a_0=1$. Models with restricted natural-parameter domains require additional admissibility conditions and are outside the regular formulation considered here.

Let $\mathcal S_K$ be the full cubic $C^2$ spline space on a partition of $[0,1]$ into $r$ intervals, with dimension $K=r+3$ and linearly independent B-spline basis $B(x)=(B_1(x),\ldots,B_K(x))^\top$. Each $g\in\mathcal S_K$ has the representation
\begin{equation}
\label{eq:spline-representation}
g_\beta(x)=B(x)^\top\beta,\qquad \beta\in\mathbb R^K.
\end{equation}
For a complete sample $(X_i,Y_i)_{i=1}^N$ with sample size $N$, we can give the log-likelihood:
\begin{equation}
\label{eq:offline-likelihood}
\ell_N(\beta)=\frac1{a_0}\sum_{i=1}^N
\left[Y_iB(X_i)^\top\beta-b\{B(X_i)^\top\beta\}\right].
\end{equation}
We use the stabilized second-derivative penalty
\begin{equation}
\label{eq:function-penalty}
J(g)=\int_0^1\{g''(x)\}^2\,dx+\epsilon\int_0^1g(x)^2\,dx,
\qquad \epsilon=10^{-8}.
\end{equation}
With Gram and penalty matrices
\begin{equation}
\label{eq:gram-penalty}
G=\int_0^1 B(x)B(x)^\top\,dx,
\qquad P=\int_0^1 B''(x)B''(x)^\top\,dx+\epsilon G,
\end{equation}
the penalty satisfies
\begin{equation}
\label{eq:penalty-matrix}
J(g_\beta)=\beta^\top P\beta.
\end{equation}
The offline estimator is
\begin{equation}
\label{eq:offline-estimator}
\widehat g_N^{\mathrm{off}}=\arg\min_{g\in\mathcal S_K}
\left\{-\frac1N\ell_N(g)+\frac{\lambda_N^{\mathrm{off}}}{2}J(g)\right\},
\end{equation}
where $\ell_N(g)$ is the log-likelihood evaluated at $g$ and $\lambda_N^{\mathrm{off}}>0$ is the normalized smoothing parameter.

\subsection{Dynamic Bases and Algorithmic Subbatches}
\label{subsec:dynamic-renewable}

As introduced in Section~\ref{sec:intro}, our motivation arise  from the initial spline space restricting representation of the estimation and resulting in bais with limited startup data. We allow the bases of the spline space to expand as data accumulate.

Let $\mathcal D_0$ denote the startup batch and $\mathcal B_r$, $r=1,2,\ldots$, the subsequent \emph{physical batches} determined by the data source. The model is updated once per \emph{algorithmic subbatch}. A physical batch is processed as one subbatch or split at basis-expansion events. We will give the detailed description of the difference between this two concepts in the following parts.

Our growth mechanism is inspired by the dynamic penalized-spline
framework of \citet{xueyao2022} and penalized-spline smoothing \citep{eilersmarx1996}. The cumulative sample size determines when the space expands, and the curvature of the preceding fitted function guides where a new knot is inserted. Each refinement introduces one interior knot, producing a sequence of nested spline spaces, 
\begin{equation}
\label{eq:nested-spaces}
\mathcal S_0
\subseteq
\mathcal S_1
\subseteq
\cdots
\subseteq
\mathcal S_j\subseteq\cdots.
\end{equation}

At each algorithmic update, the spline space at the current step either remains unchanged, so that \(\mathcal{S}_j=\mathcal{S}_{j-1}\), or undergoes a basis expansion, in which case \(\mathcal{S}_{j-1}\subsetneq\mathcal{S}_j\). Because the spline spaces are nested, every function represented in the previous space remains exactly representable after the expansion.

We fit the initial model based on the startup batch $\mathcal D_0$ with the initial spline space $\mathcal S_0$, giving $\widehat g_0$. For $j\geq1$, subbatch $\mathcal D_j$ is processed in the receiving space $\mathcal S_j$, updating the preceding fit $\widehat g_{j-1}$ to $\widehat g_j$. At an expansion, the nested-space embedding preserves the fitted function in the enlarged space and we will give the detailed transport methods in the following part.

Let $r_{\mathrm{init}}=K_0-3$ denote the initial number of
spline intervals. At cumulative sample size $t$, the finite implementation uses the target growth calendar
\begin{equation}
\label{eq:growth-calendar}
r_*(t)
=
\min\left\{
61,\,
\max\left[
r_{\mathrm{init}},
\left\lfloor
2\left(\frac{t}{200}\right)^{2/3}
\right\rfloor
\right]
\right\},
\qquad
K_*(t)=r_*(t)+3.
\end{equation}
Thus, the target dimension increases with the cumulative sample size according to the prescribed calendar. And the calendar is subject to the initial dimension and the finite implementation cap of $K_*=64$. The nominal observation time for attaining interval count $r$ is
\begin{equation}
\label{eq:growth-time}
\tau_r
=
\left\lceil
200(r/2)^{3/2}
\right\rceil.
\end{equation}
At each prescribed growth time, one new knot is inserted and the mesh-admissibility conditions are satisfied.

The cap controls the $O(K_j^2)$ storage required by the dense curvature matrix. The asymptotic analysis in Section ~\ref{sec:theory} instead uses an uncapped recursion under the growing-startup and dimension-growth conditions in Assumptions~\ref{ass:mesh} and~\ref{ass:startup}. The choice of hyperparameters here accounts for the specific computational implementation and empirical findings from previous studies.

Knot placement at update $j$ is determined by the
curvature of the pre-update fitted function
$\widehat g_{j-1}$. Let $r$ denote the current number
of spline intervals, and let
$I_k=[\kappa_{k-1},\kappa_k]$,
$k=1,\ldots,r$, denote the corresponding partition
intervals, where
$0=\kappa_0<\kappa_1<\cdots<\kappa_r=1$.
Let $\mathcal P_r=\{I_1,\ldots,I_r\}$ denote
the current partition, and define its global mesh
ratio by
\[
\rho(\mathcal P_r)
=
\frac{\max_{1\leq k\leq r}|I_k|}
     {\min_{1\leq k\leq r}|I_k|}.
\]
For each interval $I_k$, let $\mathcal P_r^{(k)}$
denote the partition obtained by bisecting $I_k$
at its midpoint. The admissible set is defined by
\[
\mathcal A_r
=
\left\{
I_k:
\rho(\mathcal P_r^{(k)})\leq 8
\right\}.
\]

We consider local windows consisting of at most seven consecutive intervals. Specifically, for
$1\le p\le q\le r$ and $q-p+1\le 7$, where $p$ and $q$ are the starting and ending interval indices of a local window, define
\[
W_{p,q}=\bigcup_{k=p}^{q}I_k.
\]

The curvature score of a window is defined as
\begin{equation}
\label{eq:knot-window-score}
S_j(W_{p,q})
=
|W_{p,q}|^2
\left\{
\int_{W_{p,q}}
[\widehat g_{j-1}''(x)]^2\,dx
\right\}^{1/2},
\end{equation}
where $|W_{p,q}|$ denotes the length of the window.

We restrict the candidate windows to those containing at least one admissible interval:
\begin{equation}
\label{eq:admissible-windows}
\mathcal W_r^{\mathrm{adm}}
=
\left\{
W_{p,q}:
\begin{array}{l}
1\le p\le q\le r,\quad q-p+1\le 7,\\
\exists k\in\{p,\ldots,q\}
\text{ such that }I_k\in\mathcal A_r
\end{array}
\right\}.
\end{equation}

The selected window is
\begin{equation}
\label{eq:selected-window}
W_j^*
=
\arg\max_{W\in\mathcal W_r^{\mathrm{adm}}}
S_j(W).
\end{equation}

\begin{equation}
\label{eq:knot-insertion}
I_j^*
=
\arg\max_{\substack{I_k\subseteq W_j^*\\I_k\in\mathcal A_r}}
|I_k|,
\qquad
x_j^{\mathrm{new}}
=
\frac{\inf I_j^*+\sup I_j^*}{2}.
\end{equation}

\begin{lemma}
If $\rho(\mathcal P_r)\leq 8$, then
$\mathcal A_r$ is nonempty.
\end{lemma}

\begin{proof}
Let $H$ and $h$ denote the largest and smallest interval lengths, respectively. Bisecting an interval of length $H$ gives a new mesh ratio bounded by
\[
\frac{H}{\min(h,H/2)}
=
\max\left\{\frac{H}{h},2\right\}
\leq 8.
\]
Thus, at least one midpoint insertion is admissible.
\end{proof}

If the maximum curvature score is below the prescribed numerical tolerance, the algorithm instead selects the globally longest admissible interval:
\begin{equation}
\label{eq:knot-fallback}
I_j^*
=
\arg\max_{I_k\in\mathcal A_r}|I_k|,
\qquad
x_j^{\mathrm{new}}
=
\frac{\inf I_j^*+\sup I_j^*}{2}.
\end{equation}
Ties are resolved by selecting the first eligible interval.

After introducing growth mechanism, we give detailed explanation for the two concepts, \emph{physical batches} and \emph{algorithmic subbatch}. The growth calendar introduces refinement boundaries in addition to physical-batch boundaries. Let $\mathcal C_{\mathrm{phys}}$ denote the set of cumulative observation indices at the ends of physical batches, and let $\mathcal C_{\mathrm{ref}}$ denote the set of cumulative observation indices immediately preceding observations at which prescribed basis-expansion events occur. The set of algorithmic cut points is defined by
\begin{equation}
\label{eq:algorithmic-boundaries}
\mathcal C_{\mathrm{alg}}
=
\mathcal C_{\mathrm{phys}}
\cup
\mathcal C_{\mathrm{ref}},
\end{equation}
with the startup endpoint $N_0$ taken as the initial cut point.
After ordering these cut points, each pair of consecutive boundaries determines an incoming algorithmic subbatch $\mathcal D_j$. Thus, a physical batch is split whenever it contains one or more basis-expansion events, whereas a physical batch without an intervening refinement event remains a single algorithmic subbatch. The receiving spline space may remain unchanged across consecutive updates.

For $j\geq 1$, let $\mathcal D_j$ denote the resulting incoming
algorithmic subbatches. Define
\begin{equation}
\label{eq:subbatch-counts}
n_0=|\mathcal D_0|,
\qquad
n_j=|\mathcal D_j|,
\qquad
N_j=n_0+\sum_{\ell=1}^{j}n_\ell.
\end{equation}
Here, $n_0$ is the startup sample size, $n_j$ is the size of the $j$th incoming algorithmic subbatch, and $N_j$ is the cumulative sample size after processing that subbatch. In particular, $N_0=n_0$. The index $j$ therefore identifies an algorithmic update. Each algorithmic subbatch is processed within a single
receiving spline space, and every basis-expansion event occurs at a subbatch boundary.

\begin{figure}[!htbp]
    \centering
    \includegraphics[width=0.82\textwidth]
    {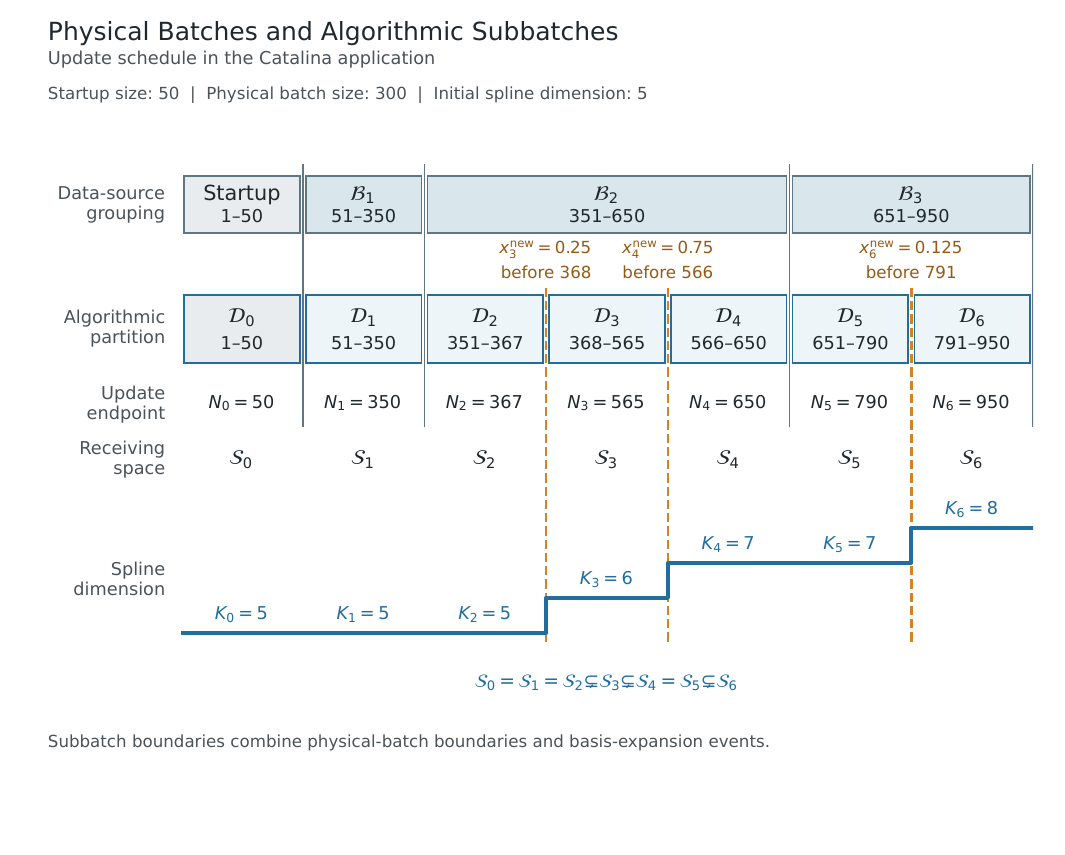}
    \caption{Construction of algorithmic subbatches from
    physical-batch boundaries and basis-refinement events.
    }
    \label{fig:batch-structure}
\end{figure}

Figure~\ref{fig:batch-structure} illustrates how the algorithmic subbatches are constructed from physical-batch boundaries and basis-refinement events. The illustration uses the the real data application to the Catalina dataset. Solid vertical lines indicate physical-batch boundaries, while dashed vertical lines indicate basis-refinement events. 

\FloatBarrier

\subsection{Renewable Estimation Algorithm for GNR}
\label{subsec:renewable-algorithm}

Our proposed algorithm begins with the startup batch $\mathcal D_0$ and an initial spline space $\mathcal S_0$ of dimension $K_0$. A penalized-likelihood fit is first obtained using the startup observations. Denote the estimated coefficient vector by $\widehat\beta_0$. The fitted function is represented by its coefficient vector in the current basis. Let $U_0(\beta)$ and $H_0(\beta)$ denote the score and negative Hessian of the startup log-likelihood respectively. At the initial fitted center, we retain
\begin{equation}
\label{eq:initial-summary}
h_0=U_0(\widehat\beta_0),
\qquad
M_0=H_0(\widehat\beta_0).
\end{equation}

After processing subbatch $\mathcal D_j$, we retain the fitted
coefficient vector $\widehat\beta_j$, the accumulated data-score
vector $h_j$, and the accumulated data-curvature matrix $M_j$.
Inspired by \citet{luosong2020}, we approximate the log-likelihood of all data processed so far by
\begin{equation}
\label{eq:historical-quadratic}
Q_j(\beta)
= h_j^\top(\beta-\widehat\beta_j)
  - \frac{1}{2}(\beta-\widehat\beta_j)^\top
    M_j(\beta-\widehat\beta_j).
\end{equation}
We will give the update steps for $(\hat\beta_j, h_j, M_j)$ in detail in the following part. 

For each incoming subbatch $\mathcal D_j$, let $B_j(x)$
denote the B-spline basis of the receiving space
$\mathcal S_j$. Its log-likelihood contribution is
\begin{equation}
\label{eq:incoming-likelihood}
\ell_j(\beta)
=
\frac{1}{a_0}
\sum_{i\in\mathcal D_j}
\left[
Y_iB_j(X_i)^\top\beta
-
b\{B_j(X_i)^\top\beta\}
\right].
\end{equation}
The corresponding score function and negative Hessian are
\begin{equation}
\label{eq:incoming-derivatives}
\begin{split}
U_j(\beta)
&=
\frac{1}{a_0}
\sum_{i\in\mathcal D_j}
B_j(X_i)
\left[
Y_i-b'\{B_j(X_i)^\top\beta\}
\right],\\
H_j(\beta)
&=
\frac{1}{a_0}
\sum_{i\in\mathcal D_j}
b''\{B_j(X_i)^\top\beta\}
B_j(X_i)B_j(X_i)^\top.
\end{split}
\end{equation}
These definitions apply to both updating cases, with
$B_j$ representing the basis of the receiving spline space.

We first consider an incoming subbatch $\mathcal D_j$ for which basis-refinement event does not happen:
\[
\mathcal S_j=\mathcal S_{j-1},
\qquad
K_j=K_{j-1}.
\]
The previous and current coefficient vectors are then expressed in the same coordinates. We can give the renewable estimator by  combining the historical quadratic approximation for likelihood, the incoming subbatch likelihood and the current penalty:
\begin{equation}
\label{eq:fixed-basis-update}
\widehat\beta_j
=
\arg\max_{\beta\in\mathbb R^{K_j}}
\left\{
Q_{j-1}(\beta)
+\ell_j(\beta)
-\frac{\lambda_{N_j}^{\mathrm{raw}}}{2}
\beta^\top P_j\beta
\right\},
\end{equation}
where the tuning details of the raw penalty is provided in the Supporting Information. 

After obtaining the fitted center $\widehat\beta_j$, the historical data-score and curvature summaries can be updated as follows:
\begin{equation}
\label{eq:fixed-basis-refresh}
\begin{split}
h_j
&=
h_{j-1}
+M_{j-1}
(\widehat\beta_{j-1}-\widehat\beta_j)
+U_j(\widehat\beta_j),\\
M_j
&=
M_{j-1}+H_j(\widehat\beta_j).
\end{split}
\end{equation}

We next consider a basis-refinement event, where
\[
\mathcal S_{j-1}\subsetneq\mathcal S_j.
\]
Let $B_{j-1}(x)$ and $B_j(x)$ denote the previous and receiving B-spline bases respectively. Since the spaces are nested, there exists a full-column-rank embedding matrix $A_j$ satisfying
\begin{equation}
\label{eq:embedding}
B_{j-1}(x)=A_j^\top B_j(x).
\end{equation}
Accordingly, the previous fitted function is preserved in the enlarged space by embedding its coefficient vector:
\begin{equation}
\label{eq:embedded-center}
\widetilde\beta_{j-1}
=
A_j\widehat\beta_{j-1}.
\end{equation}

Although this embedding preserves the fitted function, the retained historical score and curvature remain expressed in the coefficient coordinates of the previous spline space. We address this dimensional mismatch by defining a projection from the receiving space onto the embedded previous space using the Lebesgue-$L^2$ Gram geometry of the receiving basis. Let
\[
G_j=\int_0^1B_j(x)B_j(x)^\top\,dx.
\]
The corresponding Gram-metric left inverse is
\begin{equation}
\label{eq:gram-pullback}
R_j
=
(A_j^\top G_jA_j)^{-1}A_j^\top G_j.
\end{equation}
Since $A_j$ has full column rank and $G_j$ is positive
definite, this matrix is well-defined and satisfies
\begin{equation}
\label{eq:gram-left-inverse}
R_jA_j=I.
\end{equation}
Here, $I$ is the identity matrix. Moreover, $A_jR_j$ is the $G_j$-orthogonal projection onto the embedded previous coefficient space.

We extend the retained historical quadratic approximation to the receiving space by composing it with this projection:
\begin{equation}
\label{eq:historical-pullback}
Q_{j-1}^{\mathrm{PB}}(\beta)
=
Q_{j-1}(R_j\beta).
\end{equation}
The transported historical score and curvature are given:
\begin{equation}
\label{eq:transported-summaries}
h_j^-=R_j^\top h_{j-1},
\qquad
M_j^-=R_j^\top M_{j-1}R_j.
\end{equation}
By construction, the pullback preserves the retained historical quadratic approximation on the embedded previous space:
\begin{equation}
\label{eq:preservation}
Q_{j-1}^{\mathrm{PB}}(A_j\beta)
=
Q_{j-1}(\beta).
\end{equation}
Thus, the above transport method retains the historical quadratic approximation on the previous spline space and also provide a specified extension to the receiving new spline space.

The renewable estimator in the enlarged space can be give by combining the transported historical quadratic approximation with the incoming subbatch likelihood and the current penalty:
\begin{equation}
\label{eq:dynamic-basis-update}
\begin{split}
\widehat\beta_j
=
\arg\max_{\beta\in\mathbb R^{K_j}}
\Bigg\{&
(h_j^-)^\top
(\beta-\widetilde\beta_{j-1})\\
&-\frac12
(\beta-\widetilde\beta_{j-1})^\top
M_j^-
(\beta-\widetilde\beta_{j-1})\\
&+\ell_j(\beta)
-\frac{\lambda_{N_j}^{\mathrm{raw}}}{2}
\beta^\top P_j\beta
\Bigg\}.
\end{split}
\end{equation}

\begin{figure}[!htbp]
    \centering
    \includegraphics[width=0.95\textwidth]
    {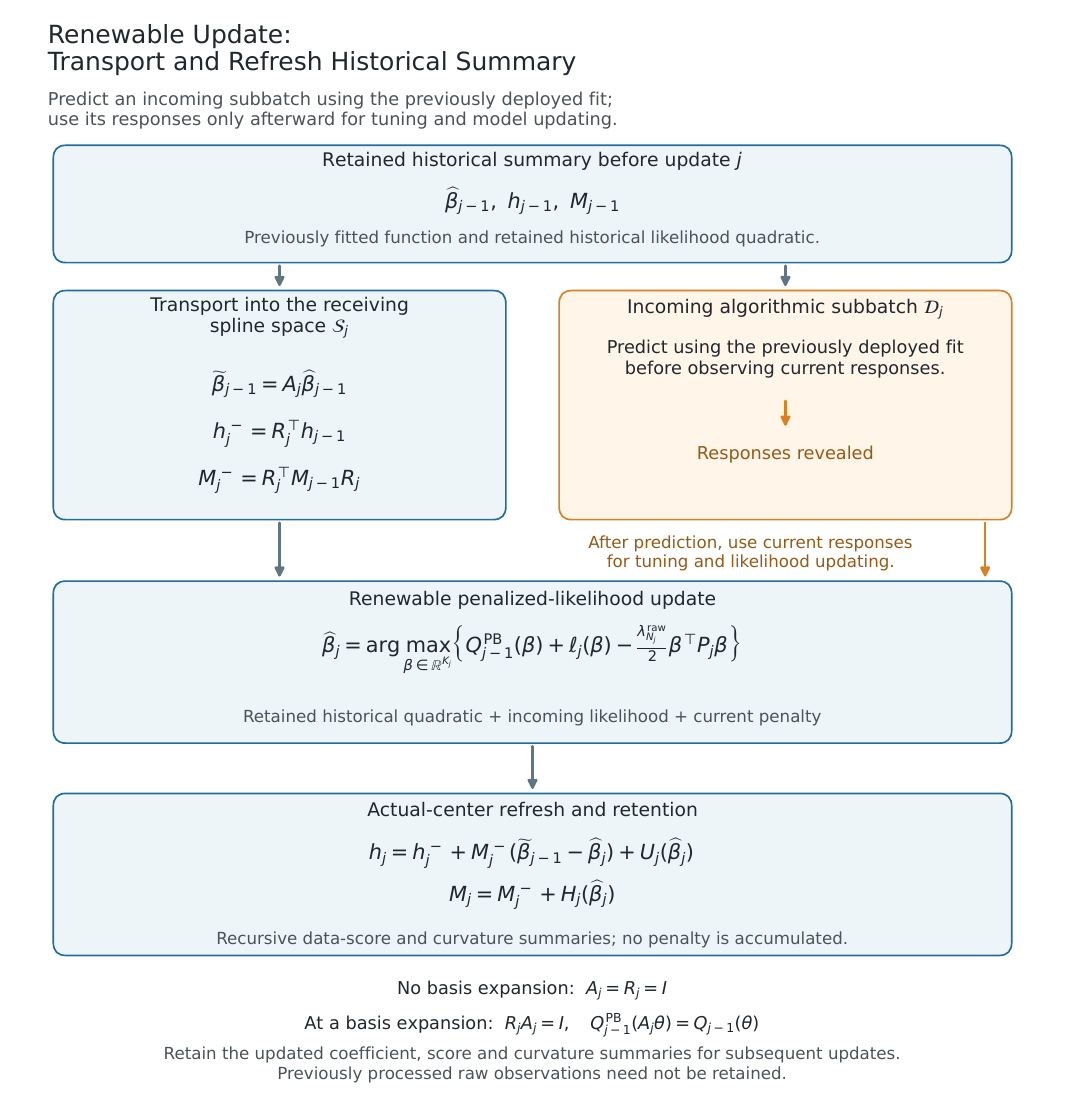}
    \caption{Illustration of the renewable algorithm}
    \label{fig:renewable-update}
\end{figure}

After solving the optimization problem, the corresponding historical summary can be updated at the actual fitted center:
\begin{equation}
\label{eq:dynamic-basis-refresh}
\begin{split}
h_j
&=
h_j^-
+M_j^-
(\widetilde\beta_{j-1}-\widehat\beta_j)
+U_j(\widehat\beta_j),\\
M_j
&=
M_j^-+H_j(\widehat\beta_j).
\end{split}
\end{equation}
When no basis expansion occurs, taking $A_j=R_j=I$ reduces the transported summaries and the dynamic-basis update to the fixed-basis recursion. Thus, both cases belong to the same renewable penalized-likelihood framework.

The smoothing parameter is selected from a prespecified finite grid. At the start of subbatch $\mathcal D_j$, put $s=N_{j-1}$. For each candidate $d$, the coefficient vector $\beta_{j,d}$ solves 
\begin{equation}
\label{eq:smoothing-candidate}
\{M_{j-1}+\lambda_d^{\mathrm{raw}}(s)P_{j-1}\}\beta_{j,d}
=M_{j-1}\widehat\beta_{j-1}+h_{j-1}.
\end{equation}
where $\lambda_d^{\mathrm{raw}}(s)$ denotes the raw penalty multiplier associated with candidate $d$. The candidate fits share the same retained historical summary and are evaluated using incoming predictive losses. An exponentially weighted score with memory $W=1000$ is used to select the current smoothing parameter. The selected multiplier is then used in the renewable penalized-likelihood update. Further details are provided in the Supporting Information.

For each incoming algorithmic subbatch $\mathcal D_j$, prequential predictions are generated using the previously estimated function $\widehat g_{j-1}$ before its responses are observed. When a basis expansion occurs, this estimated function is embedded in the receiving spline space without changing its functional representation. Once the responses become available, they may be used for smoothing-parameter selection and the subsequent renewable update, giving the updated estimation, $\widehat g_j$.

\FloatBarrier

\section{Theoretical Results}
\label{sec:theory}

Let $\mathcal H=H^2[0,1]$. For $u,v\in\mathcal H$,
extend the penalty defined in Section~\ref{sec:method}
to the bilinear form
\[
J(u,v)=\int_0^1u''(x)v''(x)\,dx
+\epsilon\int_0^1u(x)v(x)\,dx,
\]
with $J(u)=J(u,u)$. For $0<\alpha\leq1$, define
the inner product and its induced norm on $\mathcal H$ by
\begin{equation}\label{eq:alpha-inner-product}
\langle u,v\rangle_\alpha
=
\int_0^1u(x)v(x)\,dx+\alpha J(u,v),
\qquad
\|u\|_\alpha^2=\langle u,u\rangle_\alpha.
\end{equation}
We refer to this as the energy norm. For a continuous linear functional $F$ on $\mathcal H$
and a subspace $S\subseteq\mathcal H$, define
\[
\|F\|_{\alpha,*,S}
=
\sup_{\substack{u\in S\\\|u\|_\alpha\leq1}}|F[u]|.
\]
When $S=\mathcal H$, we write $\|F\|_{\alpha,*}$.

We study the uncapped, exact-arithmetic recursion in Section~\ref{subsec:renewable-algorithm}, using a triangular array with row size $N$. The penalty $J$ is defined in
\eqref{eq:function-penalty}, and
\[
\Gamma=200^{-1/5}\{10^{-6},10^{-5},\ldots,1\}.
\]
The retained summaries contain data-likelihood  contributions only. All projections use the same Lebesgue-$L^2$ inner product. Constants in the assumptions below are independent of $N$.

\begin{assumption}[Model]
\label{ass:model}
Within each row, the observations are i.i.d., with known fixed $a_0>0$, natural-parameter domain $\mathbb R$, and a fixed target $g_0\in H^2[0,1]$ common to all rows. The covariate density satisfies
\[
0<f_-\leq f_X(x)\leq f_+<\infty,
\qquad x\in[0,1].
\]
The cumulant function $b\in C^3(\mathbb R)$ satisfies
$b''(z)>0$ for every $z\in\mathbb R$.
\end{assumption}

\begin{assumption}[Spline spaces]
The startup size $m=m_N$ satisfies $m\to\infty$ and $m/N\to0$.
\label{ass:mesh}
The receiving spaces are nested full cubic $C^2$ spline spaces, with uniformly bounded global mesh ratios. For each observation $i$ in subbatch $\mathcal D_j$, let $k_i=K_j-3$ denote the number of spline intervals in $\mathcal S_j$, and let $\Pi_{[i]}$ denote the $L^2[0,1]$-orthogonal projection onto $\mathcal S_j$. There exist $c,C>0$ such that
\[
c(m\vee i)^{2/3}\leq k_i\leq C(m\vee i)^{2/3},
\qquad 1\leq i\leq N.
\]
\end{assumption}

\begin{assumption}[Startup]
\label{ass:startup}
For each $N$, the prespecified collection $\mathcal S_{0,N}$ contains at most $J_0$ uniform-knot full cubic spline spaces, with $J_0$ fixed. Every candidate $\ell$ satisfies
\[
K_{0,\ell}=r_{0,\ell}+3,\qquad
\left\lfloor2(m/200)^{2/3}\right\rfloor\leq r_{0,\ell},
\qquad K_{0,\ell}\leq Cm^{2/3}.
\]
The initial space is either selected by cross-validation from $\mathcal{S}_{0,N}$ or prescribed in advance; in the latter case, $\mathcal{S}_{0,N}$ is a singleton. The multiplier $\gamma_0$ is selected by cross-validation from $\Gamma$. Both the selected space and the multiplier are measurable functions of the startup sample. The penalized-likelihood estimator is then refitted on all startup observations, with
\[
\lambda_m^{\mathrm{off}}=\gamma_0 m^{-4/5}.
\]
\end{assumption}

\begin{assumption}[Smoothing parameters]
\label{ass:recursion}
Let $\alpha_t=t^{-4/5}$. The penalty multipliers satisfy
\[
\gamma_j\in\Gamma,\qquad
\lambda_{N_j}^{\mathrm{raw}}=\gamma_jN_j^{1/5},
\qquad j\geq0,\quad N_0=m.
\]
For $j\geq1$, $\gamma_j$ is chosen according to the prespecified smoothing-parameter selection procedure in Section~\ref{subsec:renewable-algorithm}.
\end{assumption}

\begin{assumption}[Subbatches]
\label{ass:subbatches}
Let $\mathcal F_t$ denote the $\sigma$-field representing
the information available to the algorithm after observing the first $t$ observations. For every subbatch after startup,
\[
1\leq n_j\leq b_*<\infty,
\]
and its size and receiving space are measurable with respect to $\mathcal F_{N_{j-1}}$. Conditional on $\mathcal F_{N_{j-1}}$, its observations are independent and follow the row distribution in Assumption~\ref{ass:model}.
\end{assumption}

\begin{remark*}
The design-density and likelihood-curvature conditions in Assumption~\ref{ass:model} correspond to Assumption~(GR) of \citet{huangsu2021}. Their Propositions~2.1 and~2.2 give the spline approximation and inverse bounds used under Assumption~\ref{ass:mesh}, and their Corollary~3.3 gives the offline rate for each fixed choice of startup space and smoothing multiplier in Assumption~\ref{ass:startup}. For a second-derivative penalty, the unnormalized penalty of order $N_j^{1/5}$ in Assumption~\ref{ass:recursion} agrees with Assumption~5 of \citet{xueyao2022}. The knot-growth exponent $2/3$ in Assumption~\ref{ass:mesh} is also the one used in their cubic-spline example in Section~4.1. For the present recursion, we let the startup sample grow, tune over finite candidate sets, and require each subbatch size and receiving space to be determined from past information, with uniformly bounded subbatch sizes
(Assumption~\ref{ass:subbatches}).
The local derivative bounds and the finite conditional second moment of the residuals follow from Assumption~\ref{ass:model}. The online smoothing multipliers and actual fitted centers may depend on the responses in the current subbatch; no convergence of the multipliers is required.
\end{remark*}

\begin{theorem}[Consistency and convergence rate]
\label{thm:consistency}
Under Assumptions~\ref{ass:model}--\ref{ass:subbatches}, with uncapped dimension and growing initial space, the estimator introduced above satisfies
\[
\left\|
\widehat g_N^{\mathrm{ren}}-g_0
\right\|_{\alpha_N}
=
O_p(N^{-2/5}),
\]
and
\[
\left\|
\widehat g_N^{\mathrm{ren}}-g_0
\right\|_{L^2}
=
O_p(N^{-2/5}).
\]

\end{theorem}

The result establishes $L^2$-consistency for the renewable estimator under the growing-startup and dimension-growth conditions. Since $\alpha_N=N^{-4/5}$, the energy-norm bound also gives
\[
J(\widehat g_N^{\mathrm{ren}}-g_0)=O_p(1).
\]

\begin{corollary}[Matched offline discrepancy]
\label{cor:matched-offline}
Under the same assumptions, let $\widehat g_N^{\mathrm{off,matched}}$ denote the offline estimator based on the same complete dataset, realized terminal spline space and terminal penalty multiplier as the renewable estimator. Then,
\[
\left\|
\widehat g_N^{\mathrm{ren}}
-
\widehat g_N^{\mathrm{off,matched}}
\right\|_{\alpha_N}
=
O_p\left\{
N^{-1/2}\log^8(e+N)
\right\}
=
o_p(N^{-2/5}).
\]

\end{corollary}

The proof of Theorem~\ref{thm:consistency} is given in Appendix~\ref{appendxi:proof-theorem}; the proof of Corollary~\ref{cor:matched-offline} is given in the Supporting Information. 

\begin{remark*}[Asymptotic regime and finite implementation]
The theory uses a growing startup and an uncapped spline dimension. When it comes to implementation, we conduct the simulations and applications with a limited startup size and a dimension cap.
\end{remark*}

\section{Simulation Studies}
\label{sec:simulation}

\subsection{Simulation Design}

We conducted simulations to evaluate benefits from dynamic basis expansion and Gram-metric pullback in renewable GNR. We used Bernoulli models
\[
Y_i\mid X_i=x
\sim
\operatorname{Bernoulli}
\left\{\operatorname{expit}(g_0(x))\right\},
\qquad
X_i\sim\operatorname{Uniform}(0,1).
\]

The three target true functions were
\begin{align*}
\text{S1:}\quad
g_0(x)
&=-0.5+2(x-0.5),\\
\text{S2:}\quad
g_0(x)
&=-2.5+4\exp\{-80(x-0.70)^2\},\\
\text{S3:}\quad
g_0(x)
&=1.2\sin(6\pi x)+0.6\sin(10\pi x).
\end{align*}
These functions represent a linear trend, a localized peak, and an oscillatory structure with two different frequencies.

We compared four estimators. M0 is the matched offline estimator, fitted to the complete sample using the terminal spline space and smoothing parameter realized by M3. M1 retains the startup spline space and update the model with the fixed basis. M2 uses dynamic basis expansion with naive forward transport, $h_j^-=A_jh_{j-1}$ and $M_j^-=A_jM_{j-1}A_j^\top$. M3 is the proposed method. We also report M0-CV, an offline reference independently selected by cross-validation on the complete sample. For comparability, all methods use common algorithmic subbatch boundaries determined by the prescribed growth calendar.

The primary experiment considered S1--S3, with 500 independent replications per scenario and a terminal sample size of $N=6{,}000$. The startup sample size was $n_0=100$, and the initial spline dimension was selected from
\[
K_0\in\{5,8,11,16,22,32\}
\]
using three-fold cross-validation. The physical batch size was 300. In our implementation, we both use two initial choice of dimensions of the initial spline space. Here in simulations, we use three-fold cross-validation and in Section 4.4, we give a discussion on the initial choice method about dimensions of the initial spline space.

For each replication, an independent test sample of 3,000 observations was generated. Function estimation errors were evaluated on a common grid of 1,601 points. We report integrated squared errors (ISEs) for both the natural-parameter and probability functions, together with prequential log-loss. Monte Carlo summaries are reported as means with Monte Carlo standard errors (MCSEs) in parentheses.

Additional experiments involving more complex true functions, alternative physical-batch partitions, and longer data streams are provided in the Supporting Information.

\subsection{Main Simulation Results}

Table~\ref{tab:main-simulation} reports the principal
results from primary Experiments.

\begin{table}[!htbp]
\centering
\small
\caption{Main simulation results}
\label{tab:main-simulation}
\begin{tabular}{llrrr}
\toprule
Scenario & Method
& Canonical ISE
& Probability ISE
& Prequential log-loss\\
\midrule
S1 & M0-CV
& 0.003533 (0.000180)
& 0.000161 (0.000008)
& --\\
S1 & M0
& 0.003552 (0.000177)
& 0.000161 (0.000008)
& --\\
S1 & M1
& 0.003025 (0.000131)
& 0.000136 (0.000006)
& 0.632197 (0.000211)\\
S1 & M2
& 0.005552 (0.000366)
& 0.000256 (0.000018)
& 0.632445 (0.000213)\\
S1 & M3
& 0.004040 (0.000294)
& 0.000186 (0.000014)
& 0.632317 (0.000212)\\
\midrule
S2 & M0-CV
& 0.023287 (0.000501)
& 0.000258 (0.000007)
& --\\
S2 & M0
& 0.024023 (0.000483)
& 0.000283 (0.000007)
& --\\
S2 & M1
& 0.159995 (0.012068)
& 0.002360 (0.000169)
& 0.380271 (0.000683)\\
S2 & M2
& 0.035587 (0.002385)
& 0.000390 (0.000013)
& 0.375937 (0.000447)\\
S2 & M3
& 0.024507 (0.000531)
& 0.000313 (0.000008)
& 0.375328 (0.000430)\\
\midrule
S3 & M0-CV
& 0.017103 (0.000243)
& 0.000720 (0.000010)
& --\\
S3 & M0
& 0.017235 (0.000250)
& 0.000721 (0.000010)
& --\\
S3 & M1
& 0.149873 (0.009361)
& 0.006838 (0.000444)
& 0.621758 (0.000929)\\
S3 & M2
& 0.025889 (0.000562)
& 0.001099 (0.000024)
& 0.614492 (0.000395)\\
S3 & M3
& 0.021300 (0.000435)
& 0.000903 (0.000019)
& 0.613824 (0.000370)\\
\bottomrule
\end{tabular}
\end{table}

\begin{figure}[!htbp]
\centering
\includegraphics[width=0.84\linewidth]
{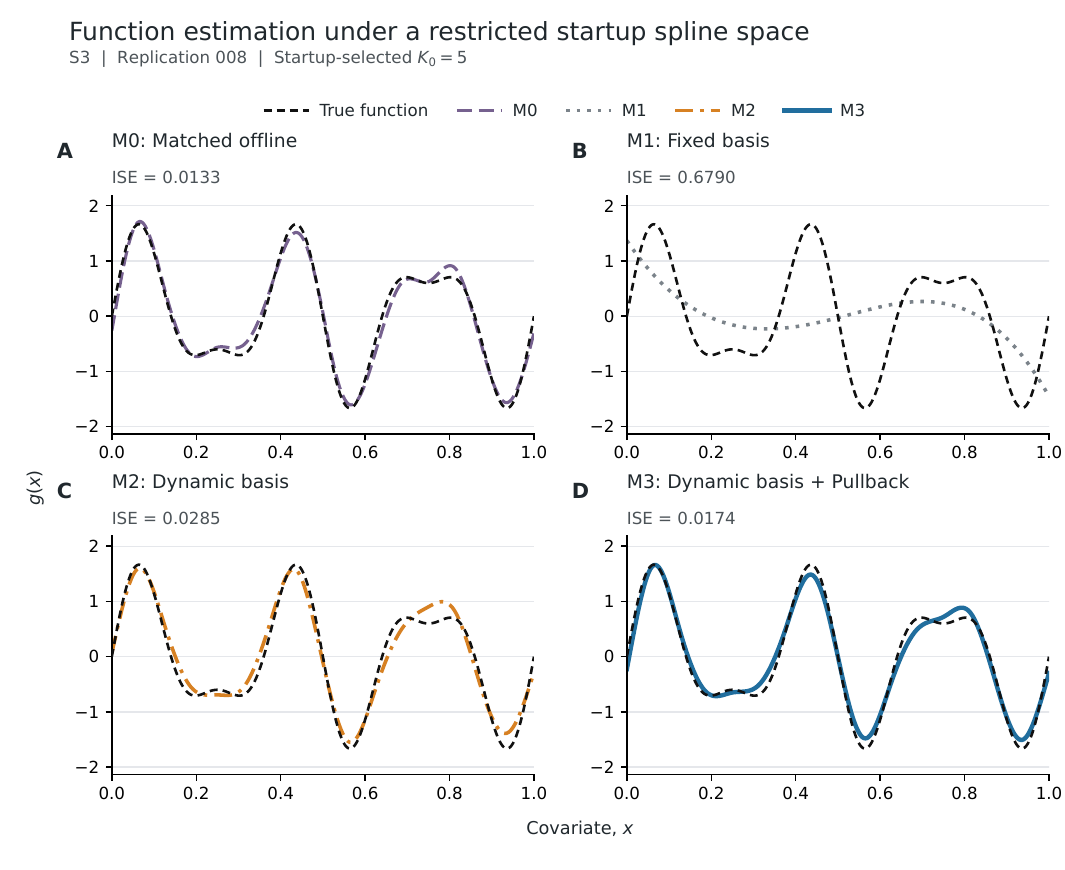}
\caption{Estimated functions for a representative
replication with a restricted startup spline space
(Scenario S3, $K_0=5$). The black dashed curve represents
the true function.}
\label{fig:representative-curves}
\end{figure}

In S1, the fixed startup space adequately represents the linear target.
For S2 and S3, both dynamic procedures substantially improve upon M1,
with M3 closest to the matched offline reference in canonical ISE.
Most of the improvement over M1 is attributable to dynamic expansion;
pullback provides a smaller additional gain.
Figure~\ref{fig:representative-curves} illustrates this pattern in an S3
replication for which startup cross-validation selected $K_0=5$.

\FloatBarrier
Table~\ref{tab:startup-selection} summarizes the initial dimensions selected
over 500 replications in S2 and S3. We classify $K_0\in\{5,8,11\}$ as
low-dimensional and $K_0\in\{16,22,32\}$ as high-dimensional.

\begin{table}[!htbp]
\centering
\small
\caption{Distribution of startup spline dimensions selected
by cross-validation over 500 replications.}
\label{tab:startup-selection}
\begin{tabular}{crrrr}
\toprule
& \multicolumn{2}{c}{S2}
& \multicolumn{2}{c}{S3}\\
\cmidrule(lr){2-3}\cmidrule(lr){4-5}
$K_0$ & Count & Proportion (\%)
& Count & Proportion (\%)\\
\midrule
5  & 81  & 16.2 & 59  & 11.8\\
8  & 47  & 9.4  & 128 & 25.6\\
11 & 83  & 16.6 & 32  & 6.4\\
16 & 71  & 14.2 & 87  & 17.4\\
22 & 83  & 16.6 & 80  & 16.0\\
32 & 135 & 27.0 & 114 & 22.8\\
\midrule
Low-dimensional
& 211 & 42.2 & 219 & 43.8\\
High-dimensional
& 289 & 57.8 & 281 & 56.2\\
\bottomrule
\end{tabular}
\end{table}

Low-dimensional spaces were selected in 42.2\% and 43.8\% of replications in S2 and S3, respectively. Thus, a restrictive initial representation can arise under the specified startup cross-validation procedure.

\begin{figure}[!htbp]
    \centering
    \includegraphics[width=0.94\linewidth]
    {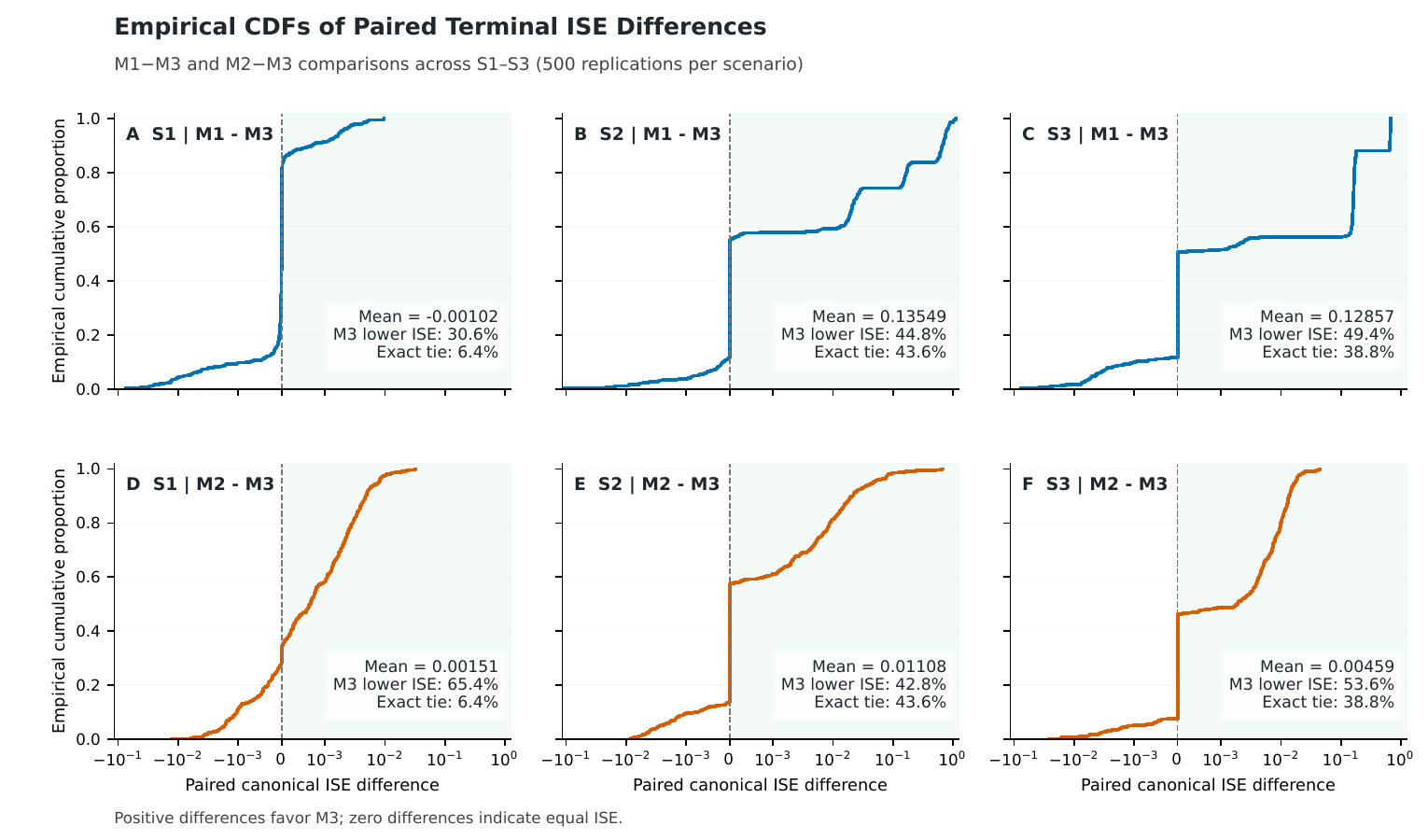}
    \caption{Empirical cumulative distribution functions of paired terminal canonical ISE differences over the 500 replications in
    each scenario.}
    \label{fig:paired-ise}
\end{figure}

To isolate the additional contribution of pullback, M2 and M3 use the same realized basis-growth path and smoothing-parameter sequence within each replication. Their paired terminal canonical ISE differences (M3 minus M2) are

\[
\begin{array}{c|rr}
\text{Scenario}
& \text{Mean difference}
& \text{MCSE}\\
\hline
\text{S1} & -0.001513 & 0.000167\\
\text{S2} & -0.011080 & 0.002292\\
\text{S3} & -0.004589 & 0.000320
\end{array}
\]
M3 has lower mean ISE in all three scenarios, although the paired results
vary across replications. In S2, M3 has lower, equal, and higher ISE than
M2 in 214, 218, and 68 replications, respectively.
Figure~\ref{fig:paired-ise} shows the empirical distributions, including
ties when no basis expansion occurs.

\FloatBarrier
\subsection{Computational Cost and Storage}

We compared renewable updating with repeated offline refitting under S2
using three paired streams, each with $N=20{,}000$, $n_0=100$,
$K_0=5$, and physical batch size 300.
Table~\ref{tab:offline-renewable} summarizes the resource requirements.

\begin{table}[!htbp]
\centering
\small
\caption{Preliminary computational comparison
between renewable updating and repeated offline
refitting ($N=20{,}000$; three paired runs).}
\label{tab:offline-renewable}
\begin{tabular}{lrr}
\toprule
& Renewable M3 & Offline\\
\midrule
Median cumulative time (s)
& 0.73 & 8.70\\
Rows supplied across updates/refits
& 19,900 & 1,048,569\\
Persistent model state / retained raw history (bytes)
& 19,520 & 240,096\\
Largest design-matrix payload (bytes)
& 108,000 & 7,360,000\\
\bottomrule
\end{tabular}
\end{table}

The median paired ratio of offline refitting time to M3 updating time was
11.48, and offline refitting processed 52.69 times as many rows. M3 also
used less persistent storage and a smaller largest design-matrix payload.
These three runs provide preliminary evidence of computational savings.

\FloatBarrier
\subsection{Initial Information and Basis Initialization}

We examined startup sizes $n_0=50$ and $n_0=200$ in S2 and S3, with
100 replications per setting. We compared fixed $K_0=5$ with joint
cross-validation over $K_0\in\{5,8,11,16,22,32\}$ and the smoothing
parameter; for fixed $K_0$, only the smoothing parameter was selected by cross-validation. Each stream contained $N=6{,}000$ observations, with an independent test sample of 3,000.

Table~\ref{tab:startup-cv} shows that $n_0=50$ often provides little
separation between candidate dimensions. With $n_0=200$, $K_0=5$ is
selected less frequently, and cross-validation more clearly favors
richer initial spaces. Thus, we take both methods for initial startup, fixed lower-dimension of the initial space and CV selection of initial dimensions, especially in our real-data applications. 

\begin{table}[!htbp]
\centering
\small
\caption{Startup cross-validation results. The table reports the selection frequency of $K_0=5$, the median validation-loss gap between the best and second-best eligible dimensions, and the mean cross-validation loss for selected candidate dimensions.}
\label{tab:startup-cv}
\begin{tabular}{llrrrrr}
\toprule
Scenario & $n_0$
& $K_0=5$ (\%)
& Median gap
& $K=5$
& $K=16$
& $K=32$\\
\midrule
S2 & 50  & 23 & $2.91\times10^{-5}$ & 0.4365 & 0.4298 & 0.4299\\
S2 & 200 & 6  & $2.04\times10^{-4}$ & 0.4195 & 0.3981 & 0.3982\\
S3 & 50  & 32 & $8.65\times10^{-5}$ & 0.6733 & 0.6627 & 0.6624\\
S3 & 200 & 2  & $9.48\times10^{-4}$ & 0.6748 & 0.6360 & 0.6367\\
\bottomrule
\end{tabular}
\end{table}

Table~\ref{tab:startup-performance} reports paired M1--M3 differences,
with positive values favoring M3. Under fixed $K_0=5$, M3 improves
terminal estimation and test performance at both startup sizes.
Under joint cross-validation, the gains decrease as $n_0$ increases,
consistent with the richer initial representations supported by more
startup information.

\begin{table}[!htbp]
\centering
\small
\caption{Paired performance differences between M1 and M3 under different startup configurations. Entries are means with paired MCSEs in parentheses over 100 replications. Positive values favor M3.}
\label{tab:startup-performance}
\begin{tabular}{llrr}
\toprule
Scenario & Initialization
& Canonical ISE difference
& Test log-loss difference\\
\midrule
S2, $n_0=50$ & Fixed $K_0=5$
& 0.7163 (0.0140) & 0.0313 (0.0004)\\
& Joint CV
& 0.1833 (0.0321) & 0.0075 (0.0012)\\
S2, $n_0=200$ & Fixed $K_0=5$
& 0.7345 (0.0139) & 0.0312 (0.0005)\\
& Joint CV
& 0.0511 (0.0175) & 0.0025 (0.0007)\\
\midrule
S3, $n_0=50$ & Fixed $K_0=5$
& 0.6589 (0.0010) & 0.0678 (0.0006)\\
& Joint CV
& 0.2633 (0.0280) & 0.0272 (0.0029)\\
S3, $n_0=200$ & Fixed $K_0=5$
& 0.6595 (0.0010) & 0.0678 (0.0006)\\
& Joint CV
& 0.0487 (0.0112) & 0.0047 (0.0012)\\
\bottomrule
\end{tabular}
\end{table}

\FloatBarrier

\section{Real-Data Applications}
\label{sec:applications}

Both applications used startup $n_0=50$ and $K_0=5$, with the initial smoothing parameter selected from startup data. Subsequent selection follows Section~\ref{subsec:renewable-algorithm}. The Supporting Information reports Catalina results with $n_0=100$ and startup cross-validation. We compare M0--M3 which is defined in Section~\ref{sec:simulation}. And we report terminal distances to M0, prequential metrics, and reserve predictions.

\subsection{Catalina: Periodic Variable Star Classification}
\label{subsec:catalina-scalar}

We first considered the Southern Catalina periodic variable-star catalogue of \citet{drake2017catalina}. The response distinguishes the selected pulsating-star classes from eclipsing binaries, and the scalar covariate is
\[
X=\log_{10}\{\text{period (days)}\}.
\]
After applying the prespecified quality and period filters, 33,648 observations were retained. Five-degree sky blocks were used to construct the training, internal validation, and reserve sets, containing 19,129, 8,409, and 6,110 observations respectively.
Because the catalogue does not provide a natural observation-time ordering, we constructed two pseudo-stream orderings based on the
record identifiers. Order 1 is used for the primary analysis, while Order 2 provides an alternative ordering for sensitivity assessment.

The first 50 training observations were used for initialization. Starting from $K_0=5$, the procedure accepted 39 basis expansions over 103 updates, reaching a terminal dimension of $K=44$. All 19,079 post-startup observations contributed to the prequential evaluation.

Table~\ref{tab:catalina-prediction} shows substantial gains from dynamic expansion over M1. The additional prequential improvement from pullback is small: relative to M2, M3 reduces log-loss by 0.001465 and increases AUC by 0.001627. M3 also improves the three validation point metrics. The two dynamic procedures have nearly identical reserve log-loss and AUC, and the paired sky-block intervals for their reserve contrasts include zero.

\begin{table}[htbp]
\centering
\footnotesize
\caption{Catalina Order 1: prequential and terminal
prediction results. The reserve was examined in
previous analyses and is reported descriptively.}
\label{tab:catalina-prediction}
\begin{tabular}{llrrr}
\toprule
Evaluation & Method & Log-loss & Brier & AUC\\
\midrule
Prequential & M1 & 0.559524 & 0.187561 & 0.655695\\
Prequential & M2 & 0.496775 & 0.163748 & 0.763534\\
Prequential & M3 & 0.495310 & 0.163280 & 0.765161\\
\midrule
Validation & M0 & 0.482799 & 0.159353 & 0.781291\\
Validation & M1 & 0.561765 & 0.188467 & 0.652077\\
Validation & M2 & 0.486071 & 0.160147 & 0.777566\\
Validation & M3 & 0.484354 & 0.159678 & 0.779445\\
\midrule
Reserve & M0 & 0.529092 & 0.178252 & 0.769190\\
Reserve & M1 & 0.602793 & 0.206273 & 0.656023\\
Reserve & M2 & 0.528801 & 0.178623 & 0.767898\\
Reserve & M3 & 0.528794 & 0.178255 & 0.767912\\
\bottomrule
\end{tabular}
\end{table}

\begin{figure}[!htbp]
    \centering
    \includegraphics[width=0.88\textwidth]
    {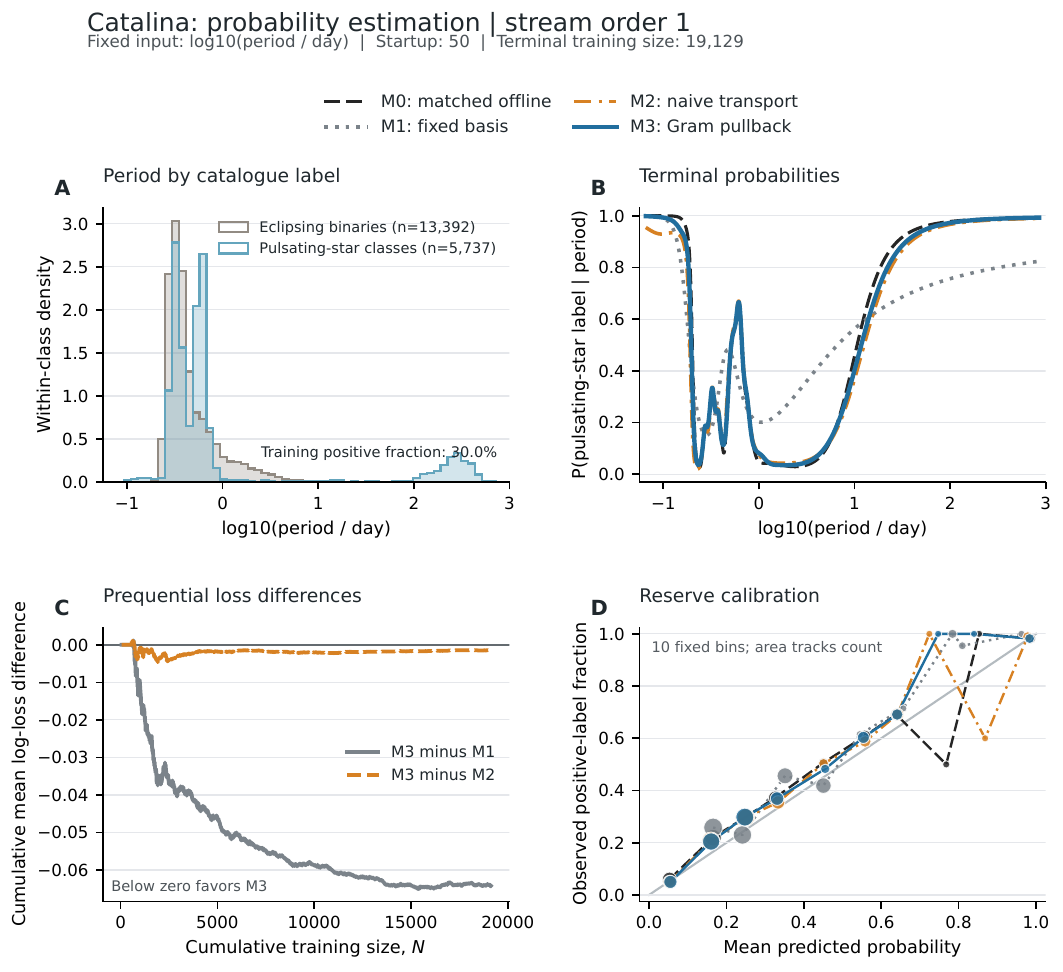}
    \caption{Catalina Order 1 application. Panel (a) shows
    the training distributions of log-period by catalogue
    label; panel (b) shows the terminal probability estimates
    from M0--M3; panel (c) shows cumulative mean prequential
    log-loss differences for M3 relative to M1 and M2, with
    values below zero favoring M3; and panel (d) shows a
    descriptive ten-bin calibration plot on the previously
    examined reserve, with point areas proportional to bin
    counts.}
    \label{fig:catalina-application}
\end{figure}

The empirically weighted probability RMS distances to matched M0 are
0.162728, 0.021398, and 0.015479 for M1, M2, and M3, respectively;
the corresponding logit RMS distances are 1.176308, 0.287125, and
0.204056. Dynamic refinement thus improves agreement with M0, with
a further reduction from pullback. Figure~\ref{fig:catalina-application}
summarizes the fitted functions and predictive results.

\FloatBarrier

\subsection{Voice: Classification from Modal Frequency}
We used the ``Voice Gender'' dataset \citep{becker2016gender}, containing
3,168 recording-level observations and 20 acoustic summaries. The scalar
covariate is modal frequency, \texttt{mode}, with female voices coded
as 1. The training and reserve sets contain 2,534 and 634 observations,
respectively. Identical acoustic-feature records were grouped during
splitting, and observations with zero modal frequency were retained.
The first 50 training observations initialized the fit with $K_0=5$.
Eight accepted expansions increased the dimension to $K=13$, with
17 updates and 2,484 prequential predictions.

Table~\ref{tab:voice-prediction} reports the prediction results. Relative
to M2, M3 reduces prequential log-loss by 0.002095 and Brier score by
0.001373, while increasing AUC by 0.007144. M3 also improves all three
point metrics on the previously examined reserve.

\begin{table}[htbp]
\centering
\footnotesize
\caption{Voice: prequential and reserve prediction results.
The reserve contains previously examined observations.}
\label{tab:voice-prediction}
\begin{tabular}{llrrr}
\toprule
Evaluation & Method & Log-loss & Brier & AUC\\
\midrule
Prequential & M1 & 0.595057 & 0.204705 & 0.738291\\
Prequential & M2 & 0.593566 & 0.205143 & 0.735980\\
Prequential & M3 & 0.591471 & 0.203770 & 0.743124\\
\midrule
Reserve & M0 & 0.551018 & 0.187142 & 0.783842\\
Reserve & M1 & 0.582703 & 0.198748 & 0.759441\\
Reserve & M2 & 0.571941 & 0.195219 & 0.769154\\
Reserve & M3 & 0.565875 & 0.192949 & 0.772428\\
\bottomrule
\end{tabular}
\end{table}

\begin{figure}[!htbp]
    \centering
    \includegraphics[width=0.88\textwidth]
    {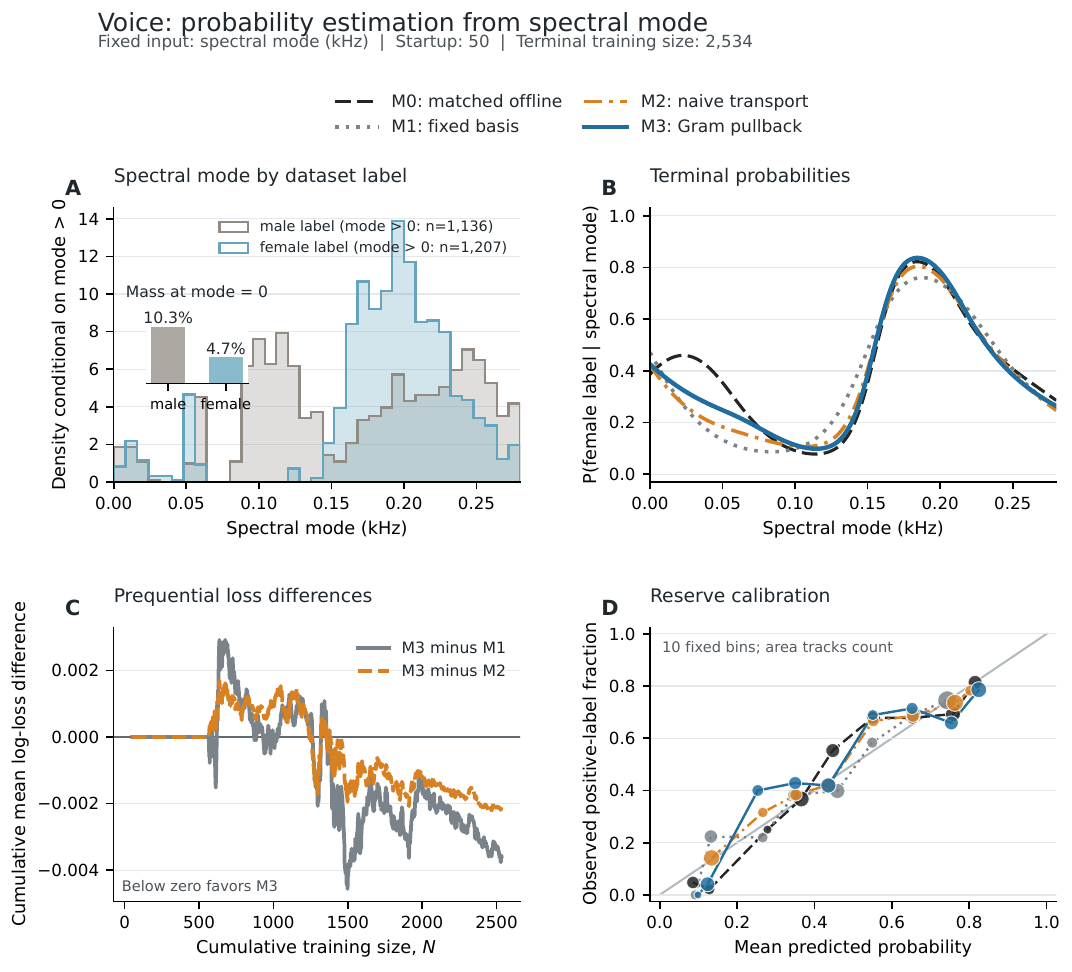}
    \caption{Voice application using spectral mode as the
    scalar covariate. Panel (a) shows the training
    distributions by label and the mass at mode zero; panel
    (b) shows the terminal probability estimates from M0--M3;
    panel (c) shows cumulative mean prequential log-loss
    differences for M3 relative to M1 and M2, with values
    below zero favoring M3; and panel (d) shows descriptive
    ten-bin calibration on the previously examined reserve,
    with point areas proportional to bin counts.}
    \label{fig:voice-application}
\end{figure}

\begin{samepage}
The probability RMS distances to matched M0 are 0.071056, 0.044234,
and 0.031994 for M1, M2, and M3, respectively; the corresponding logit
RMS distances are 0.417604, 0.245655, and 0.169442. M3 reduces the
probability RMS distance by approximately 27.7\% relative to M2.
Figure~\ref{fig:voice-application} summarizes the application.
\par
\end{samepage}

\FloatBarrier

\subsection{Comparison with External Online Classifiers}

We compared GNR with four online classifiers combining logistic or hinge
loss with either the standardized scalar input or 64 random Fourier
features (RFF64). The classifiers use stochastic gradient descent
\citep{bottou2010large}; RFF64 approximates a Gaussian kernel
\citep{rahimi2007random}, and hinge-loss probabilities use Platt-type
sigmoid calibration \citep{platt1999probabilistic}.

All methods use the same scalar inputs and stream orderings, with each
block predicted before its responses are revealed. External tuning
parameters are fixed after startup, whereas GNR retains online
smoothing-parameter selection. M3 has the highest prequential AUC in
the three comparisons in Table~\ref{tab:external-auc}; these comparisons
use the stated tuning protocols.

\begin{table}[htbp]
\centering
\small
\caption{Prequential AUC comparison with external online
classifiers under the startup-fixed tuning protocol.}
\label{tab:external-auc}
\begin{tabular}{lccc}
\toprule
Method & Catalina O1 & Catalina O2 & Voice\\
\midrule
M1 & 0.6557 & 0.6361 & 0.7383\\
M2 & 0.7635 & 0.7629 & 0.7360\\
\textbf{M3} & \textbf{0.7652} & \textbf{0.7653}
& \textbf{0.7431}\\
\midrule
Linear SGD Logistic & 0.6155 & 0.5999 & 0.5258\\
Linear SGD Hinge SVM & 0.5316 & 0.5069 & 0.5053\\
RFF64 SGD Logistic & 0.6140 & 0.6011 & 0.6711\\
RFF64 SGD Hinge SVM & 0.6577 & 0.4972 & 0.5204\\
\bottomrule
\end{tabular}

\vspace{2pt}
\begin{minipage}{0.96\textwidth}
\footnotesize
AUC is computed from the probability predictions.
For hinge-loss classifiers, probabilities are obtained
using the online sigmoid calibration procedure.
\end{minipage}
\end{table}

\FloatBarrier

\subsection{Supplementary Application and Data Availability}
HTRU2 is reported in the Supporting Information as an additional
pulsar-classification example. Across the four methods and three stream
orders, terminal holdout AUCs ranged from 0.97291 to 0.97377.

The Catalina catalogue is available from the original
article by \citet{drake2017catalina}, together with its
supplementary data:
\href{https://authors.library.caltech.edu/records/tj2s4-amz72}
{The Catalina Surveys Southern Periodic Variable Star Catalogue}.
The Voice dataset is available from Kory Becker's
original dataset repository:
\href{https://www.kaggle.com/datasets/primaryobjects/voicegender}
{Gender Recognition by Voice}.

\subsection{An Exploratory Extension to a Two-Component Logistic GAM}
\label{subsec:catalina-gam}

For the Catalina catalogue labels in Section~\ref{subsec:catalina-scalar} ($Y_i=1$ for selected pulsating stars and $Y_i=0$ for eclipsing binaries), we fitted the additive model
\[
  \operatorname{logit}\Pr(Y_i=1\mid X_i)
    = \beta_0 +  f_1(X_{i1})+f_2(X_{i2}),\qquad
  X_{i1}=\log_{10}(\text{period in days}),\quad X_{i2}=\texttt{Vamp}.
\]
Period gives the timescale of variation. The catalogue field \texttt{Vamp} gives the amplitude of the light curve fitted by adaptive Fourier decomposition, in magnitudes. We want to give an exploratory extension to 2 covariates model fitting.

We used two cubic B-spline terms with zero integral on their respective startup-transformed coordinates and an intercept. At each scheduled growth time (Section~\ref{subsec:dynamic-renewable}), we inserted one knot in the component with the higher curvature score. Let $\theta=(\beta_0,\theta_1^\top,\theta_2^\top)^\top$ collect the
intercept and the coefficients of the two centered spline terms.
For component $k$, let $R_{jk}$ be the Gram-metric left inverse
in \eqref{eq:gram-pullback}, with $R_{jk}=I$ when its knots are unchanged.
M3 transports the joint historical score and curvature as
\[
R_j=\operatorname{diag}(1,R_{j1},R_{j2}),\qquad
h_j^-=R_j^\top h_{j-1},\qquad
M_j^-=R_j^\top M_{j-1}R_j.
\]
Let $z_{ij}$ be the joint basis vector for observation $i$ in the
receiving space, so that $\eta_{ij}=z_{ij}^\top\theta$.
The two components are updated jointly by
\[
\widehat\theta_j=\arg\max_\theta
\left\{
  \sum_{i\in\mathcal D_j}
  \bigl[Y_i\eta_{ij}-\log(1+e^{\eta_{ij}})\bigr]
  +Q_{j-1}(R_j\theta)
  -\frac{\lambda_{N_j}^{\mathrm{raw}}}{2}\theta^\top P_j\theta
\right\},
\]
where $Q_{j-1}$ is the retained joint quadratic, including its
cross-component terms, and $P_j$ is the joint penalty matrix.
The summaries are refreshed at $\widehat\theta_j$ using
\eqref{eq:dynamic-basis-refresh}.
M2 used forward transport with M3's knots, penalties and subbatches.

Table~\ref{tab:catalina-gam-preq} reports predictions for two orderings of the same catalogue objects, with $n_0=50$. In both orderings, the largest improvement was from the fixed-basis M1 to the dynamic methods. Switching from M2 to M3 reduced log-loss by less than 0.002 in each order.

\begin{table}[!htbp]
\centering
\footnotesize
\caption{Catalina two-component GAM. Prequential scores for 19,079
objects per order after startup ($n_0=50$).}
\label{tab:catalina-gam-preq}
\begin{tabular}{lrrr@{\hspace{1.6em}}rrr}
\toprule
& \multicolumn{3}{c}{Order 1} & \multicolumn{3}{c}{Order 2}\\
\cmidrule(lr){2-4}\cmidrule(lr){5-7}
Method & Log-loss & Brier & AUC & Log-loss & Brier & AUC\\
\midrule
M1 & 0.511453 & 0.169218 & 0.746447 & 0.518035 & 0.171550 & 0.745473\\
M2 & 0.451191 & 0.146435 & 0.819462 & 0.458235 & 0.148848 & 0.818711\\
M3 & 0.449248 & 0.145782 & 0.823184 & 0.456500 & 0.148145 & 0.821095\\
\bottomrule
\end{tabular}
\end{table}

Of the 39 new knots, period received 24 and amplitude 15 in Order~1,
compared with 26 and 13 in Order~2. These counts describe how the fitted curvature scores allocated knots along each stream. They neither rank variable importance nor establish an optimal allocation.

The extension illustrates how a catalogue classifier could update its period and amplitude curves as objects are added. The pseudo-streams use the catalogue labels and order objects by identifier. Neither follows an observation-time sequence. 

\FloatBarrier

\section{Conclusion and Discussion}
\label{sec:discussion}
We developed a renewable estimation algorithm for nonparametric regression under canonical exponential-family, combining likelihood-based estimation, penalized splines, and dynamic basis expansion. When the spline space expands, Gram-metric pullback provides a mechanism for representing the retained historical summaries in the new spline space. The pullback has useful algebraic properties, preserving the historical quadratic in the embedded previous spline space across basis expansions. Simulation studies and real-data applications demonstrate the improvement from dynamic basis expansion when the initial spline representation is insufficient, with additional gains from Gram-metric pullback.

When the initial spline space is sufficiently rich, the improvement from dynamic basis expansion becomes relatively limited, and the additional gains from pullback are generally smaller. In such cases, the use of pullback is primarily motivated by its favorable algebraic properties for transporting historical summaries. The Gram-metric left inverse provides a geometrically specified extension based on the Lebesgue-$L^2$ structure of the receiving spline space. This gives the historical summary a well-defined algebraic representation after basis expansion, while leaving newly introduced directions to be informed by subsequent data. The pullback therefore primarily serves as a structured historical-summary transport mechanism. Its additional contribution to estimation and prediction performance is secondary. 

Several directions remain for future research. These include establishing further asymptotic distributional results, such as functional central limit theorems and Brownian-bridge-type limits, extending the framework to multivariate nonparametric regression such as additive models, or investigating practical dimension-reduction approaches such as principal component analysis (PCA). The Catalina extension in Section~\ref{subsec:catalina-gam} illustrates joint updating in a two-component model. A general asymptotic theory for renewable additive models would need to account for knot allocation across components and transport of the joint historical quadratic. Another direction is to explore alternative nonparametric representations, including deep neural networks (DNNs), within a renewable estimation framework, with particular attention to the representation of historical information and the development of corresponding theoretical guarantees.

\section{Declaration of Generative AI}
We use ChatGPT and DeepL to improve our English expression and to assist with organizing the article content and editing in our study.

\FloatBarrier
\bibliographystyle{plainnat}
\bibliography{references}

\appendix
\section[Proof of Theorem 3.1]{Proof of Theorem~\ref{thm:consistency}}
\label{appendxi:proof-theorem}
We prove Theorem~\ref{thm:consistency}. The proof of Corollary~\ref{cor:matched-offline} is given in the Supporting Information.

The main idea in this proof is as follows: decompose $\widehat g_N^{\mathrm{ren}}-g_0
=(\widehat g_N^{\mathrm{ren}}-s_N)+(s_N-g_0)$, where
$s_N=\Pi_Jg_0$. The second term is controlled by classical spline approximation. For the statistical term, we first establish stochastic bounds and startup rates and then we analyze renewable curvature, gradients and accumulated error. Write $\|\cdot\|_2=\|\cdot\|_{L^2}$. 
\subsection{Approximation for Spline Space and Stochastic Bounds}
\label{app:approx-stochastic}

By Assumptions~\ref{ass:model}--\ref{ass:mesh}, Proposition~2.1 of \citet{huangsu2021} and standard quasi-uniform spline projection stability, for $s_t=\Pi_jg_0$, $t=N_j$, we have $K_j\asymp t^{2/3}$ and
\[
 \|s_t-g_0\|_2=O(t^{-4/3}),\quad
 \|s_t-g_0\|_\infty=O(t^{-1}),\quad
 \|s_t-g_0\|_{\alpha_t}=O(t^{-2/5}).
\]
Here $J(s_t)=O(1)$; moreover,
$\|\Pi_j u\|_\alpha\le C\|u\|_\alpha$ and
$\|u\|_\infty\le C\alpha^{-1/8}\|u\|_\alpha$
for $u\in H^2[0,1]$ and $0<\alpha\le1$.

For $\xi_i=Y_i-b'\{g_0(X_i)\}$ and $\Pi_{[i]}=\Pi_j$
when $i\in\mathcal D_j$, define, for $u,v\in H^2[0,1]$,
\begin{align*}
 Z_t[u]
 &=\frac1t\sum_{i\le t}\xi_i(\Pi_{[i]}u)(X_i),\\
 V_t[u,v]
 &=\frac1t\sum_{i\le t}
 \left[
   (\Pi_{[i]}u)(X_i)(\Pi_{[i]}v)(X_i)
   -P_X\{(\Pi_{[i]}u)(\Pi_{[i]}v)\}
 \right].
\end{align*}
$P_X$ integrates a fresh covariate with the fixed realized projector. All functional and operator norms use the $\alpha$
inner product on $H^2$.
Based on Riesz representation, martingale orthogonality and the rank-one Hilbert--Schmidt identity, for deterministic $t\ge m$, we give
\[
 \E\|Z_t\|_{\alpha_t,*}^2=O(t^{-4/5}),\qquad
 \E\|V_t\|_{\mathrm{HS},\alpha_t}^2=O(t^{-3/5}).
\]
The post-startup increments are martingale differences, since
the received projection precedes the data from the subbatch. The startup bound is then determined by summing the square norm bounds of a fixed, finite set of deterministic candidates prior to selection. Fixing the norm to binary blocks, based on Doob's $L^2$ maximum inequality and the Markov inequality, the total failure probability is given by
$O((1+\log m)^{-7})=o(1)$ for the simultaneous bounds,
$m\le t\le N$,
\[
 \|Z_t\|_{\alpha_t,*}\le Ct^{-2/5}L(t),\qquad
 \|V_t\|_{\mathrm{op},\alpha_t}\le Ct^{-3/10}L(t),
 \quad L(t)=\log^4(e+t).
\]
The identity-projection versions $Z_t^I,V_t^I$ satisfy the same bounds.

\subsection{Startup Estimation and Initial Error}
\label{app:startup}

Under Assumptions~\ref{ass:model}--\ref{ass:recursion}, startup estimation is an offline penalized cubic-spline likelihood problem with $m\to\infty$, $K_{0,\ell}\asymp m^{2/3}$ and $\lambda_{m,\gamma}\asymp m^{-4/5}$. These satisfy the conditions of Theorems~3.1--3.2 and Corollary~3.3 of \citet{huangsu2021}. Their argument also extends to fixed stabilizations in $J$: The $J$ of the approximating splines is bounded, the penalty norm remains uniformly equivalent, and the added quadratic term preserves the curvature and scoring bounds. Combined with the Sobolev evaluation inequality, these results allow for selection from a fixed finite set of candidate functions and precise refitting,
\[
 \|\widehat g_0-g_0\|_{\alpha_m}=O_p(m^{-2/5}),
 \qquad
 \|\widehat g_0-g_0\|_\infty=O_p(m^{-3/10}).
\]
For $e_0=\widehat g_0-g_0$ and
$C_0=\sum_{i=1}^m e_0(X_i)^2$, the identity-projection covariance bound in Section~\ref{app:approx-stochastic} applies directly to the data-dependent $e_0$, yielding
\[
 \frac{C_0}{m}
 \le f_+\|e_0\|_2^2
 +\|V_m^I\|_{\mathrm{op},\alpha_m}\|e_0\|_{\alpha_m}^2
 =O_p(m^{-4/5}),
 \qquad
 {C_0=O_p(m^{1/5}).}
\]
This argument requires no independence between the startup
estimator and its observations.

\subsection{Curvature and Gradient Control}
\label{app:curvature-gradient}

Denote $s=N_{j-1}$, so $t-s=n_j\le b_*$. Let $\mathcal T_\ell$ be the quadratic Taylor polynomial of the negative subbatch log-likelihood $-\ell_\ell$ at its actual fitted center $\widehat g_\ell$.
Gram pullback transports the historical quadratic summary to the enlarged spline space by projecting its argument onto the previous space. Once we have the current estimation, the actual-center update adds the quadratic Taylor approximation of the incoming negative log-likelihood. Since $\Pi_\ell\Pi_{j-1}=\Pi_\ell$ for $\ell<j$, up to an additive constant, we have:
\begin{equation}\label{eq:app-terminal-objective}
 F_t^{\mathrm{ren}}(g)
 =\frac1t\left\{\sum_{\ell<j}\mathcal T_\ell(\Pi_\ell g)-\ell_j(g)\right\}
 +\frac{\gamma_j\alpha_t}{2}J(g),\qquad g\in\mathcal S_j.
\end{equation}
Its positive stabilization makes this finite-dimensional convex criterion coercive and strictly convex. In differentiation, the realized multiplier $\gamma_j$ is held fixed.

To obtain curvature uniformly in the current $g$, we first assume
$\max_{\ell<j}\|\widehat g_\ell-g_0\|_\infty\le1$.
For $i\in\mathcal D_\ell$, $i\le s$, put $z_i=\widehat g_\ell(X_i)$. Differentiation gives
\begin{align*}
 D^2F_t^{\mathrm{ren}}(g)[u,u]
 ={}&\frac1{a_0t}\sum_{i\le s}b''(z_i)\{(\Pi_{[i]}u)(X_i)\}^2\\
 &+\frac1{a_0t}\sum_{s<i\le t}b''(g(X_i))u(X_i)^2
 +\gamma_j\alpha_tJ(u),\qquad u\in\mathcal S_j.
\end{align*}
Assumption~\ref{ass:model} bounds $a_0^{-1}b''(z_i)$ below by a fixed $w_*>0$; the incoming term is nonnegative. With $\gamma_{\min}=\min\Gamma>0$,
\[
 D^2F_t^{\mathrm{ren}}(g)[u,u]
 \ge \frac{w_*}{t}\sum_{i\le s}\{(\Pi_{[i]}u)(X_i)\}^2
 +\gamma_{\min}\alpha_tJ(u).
\]
Apply the results in Section~\ref{app:approx-stochastic} to the complete historical prefix:
\[
 \frac1t\sum_{i\le s}\{(\Pi_{[i]}u)(X_i)\}^2
 =\frac1t\sum_{i\le s}P_X\{(\Pi_{[i]}u)^2\}
 +\frac{s}{t}V_s[u,u].
\]
On its uniform event, $s=t-O(1)$ implies
$|(s/t)V_s[u,u]|\le Ct^{-3/10}L(t)\|u\|_{\alpha_t}^2$.
To find the lower bound of the nonnegative population sum, retain
$\lceil t/2\rceil\le i\le s$. There are at least $t/4$ such observations for large $m$, and $t/2\le m\vee i\le t$ gives $k_i\asymp t^{2/3}$ by Assumption~\ref{ass:mesh}. The density bound, the Lebesgue-$L^2$ Pythagorean identity and spline approximation give
\begin{align*}
 P_X\{(\Pi_{[i]}u)^2\}
 &\ge f_-\|\Pi_{[i]}u\|_2^2
 =f_-\{\|u\|_2^2-\|(I-\Pi_{[i]})u\|_2^2\}\\
 &\ge f_-\|u\|_2^2-Ck_i^{-4}J(u)
 \ge f_-\|u\|_2^2-Ct^{-8/3}J(u).
\end{align*}
This restriction is a lower-bound argument and the algorithm preserves all historical summaries. Consequently,
\[
 D^2F_t^{\mathrm{ren}}(g)[u,u]
 \ge c\|u\|_2^2+(\gamma_{\min}\alpha_t-Ct^{-8/3})J(u)
 -Ct^{-3/10}L(t)\|u\|_{\alpha_t}^2.
\]
Since $t^{-8/3}=o(\alpha_t)$ and $t^{-3/10}L(t)=o(1)$ uniformly for $t\ge m\to\infty$, there exists a fixed $c_*>0$ such that
\begin{equation}\label{eq:app-strong-convexity}
 D^2F_t^{\mathrm{ren}}(g)[u,u]\ge c_*\|u\|_{\alpha_t}^2,
 \qquad g,u\in\mathcal S_j.
\end{equation}

Control the gradient at $s_t=\Pi_jg_0$.
Put $\mu=b'$ and $v_i=(\Pi_{[i]}g_0)(X_i)$ for all $i\le t$. For $i\le s$, define
\[
 \rho_i=\mu(z_i)+\mu'(z_i)(v_i-z_i)-\mu(v_i).
\]
Use $\Pi_\ell\Pi_j=\Pi_\ell$ in \eqref{eq:app-terminal-objective}, for $u\in\mathcal S_j$ we have,
\begin{align}\label{eq:app-gradient-decomposition}
 DF_t^{\mathrm{ren}}(s_t)[u]
 ={}&-a_0^{-1}Z_t[u]+\gamma_j\alpha_tJ(s_t,u)\nonumber\\
 &+\frac1{a_0t}\sum_{i\le t}[\mu(v_i)-\mu(g_0(X_i))](\Pi_{[i]}u)(X_i)\nonumber\\
 &+\frac1{a_0t}\sum_{i\le s}\rho_i(\Pi_{[i]}u)(X_i).
\end{align}
Here $J(\cdot,\cdot)$ is the bilinear form associated with $J$.
These terms are decomposed into the score, current penalty, representation bias and historical Taylor remainder. The exact incoming likelihood based on incoming subbatch contributes no Taylor remainder.
Section~\ref{app:approx-stochastic} and Cauchy--Schwarz give
\[
 |(\Pi_{[i]}u)(X_i)|\le Ct^{1/10}\|u\|_{\alpha_t},\qquad
 |\gamma_j\alpha_tJ(s_t,u)|\le Ct^{-2/5}\|u\|_{\alpha_t}.
\]
Assumption~\ref{ass:mesh} and integral comparison give
\[
 \sum_{i\le t}k_i^{-3/2}\le C\{1+\log(t/m)\},\qquad
 \sum_{i\le t}k_i^{-3}\le Cm^{-1}.
\]
Local boundedness of $\mu'$ and the spline sup-norm approximation bound therefore give representation-bias dual norm at most
$Ct^{-9/10}\{1+\log(t/m)\}$.
Historical localization and the bound on $b'''$ give
$|\rho_i|\le C|z_i-v_i|^2$ by Taylor's theorem, and thus
\[
 \sum_{i\le s}|\rho_i|
 \le C\left\{C_{j-1}+\sum_{i\le s}k_i^{-3}\right\}
 \le C(C_{j-1}+m^{-1}),\qquad
 C_r:=\sum_{\ell=0}^{r}\sum_{i\in\mathcal D_\ell}
 \{\widehat g_\ell(X_i)-g_0(X_i)\}^2.
\]
Combining the four bounds yields
\begin{equation}\label{eq:app-gradient-bound}
 \|DF_t^{\mathrm{ren}}(s_t)\|_{\alpha_t,*,\mathcal S_j}
 \le C\left\{\|Z_t\|_{\alpha_t,*}+t^{-2/5}
 +t^{-9/10}[C_{j-1}+1+\log(t/m)]\right\}.
\end{equation}
For $d=\widehat g_j-s_t$, exact stationarity and
\eqref{eq:app-strong-convexity} imply the usual gradient-to-error bound:
\[
 c_*\|d\|_{\alpha_t}^2
 \le [DF_t^{\mathrm{ren}}(\widehat g_j)-DF_t^{\mathrm{ren}}(s_t)][d]
 =-DF_t^{\mathrm{ren}}(s_t)[d].
\]
Adding the approximation error and applying the uniform score and evaluation bounds gives
\begin{align}
 \|\widehat g_j-g_0\|_{\alpha_t}
 &\le C\{t^{-2/5}L(t)+t^{-9/10}C_{j-1}\},\label{eq:app-energy-recursion}\\
 \|\widehat g_j-g_0\|_\infty
 &\le C\{t^{-3/10}L(t)+t^{-4/5}C_{j-1}\}.\label{eq:app-sup-recursion}
\end{align}

\subsection{Recursive Error Control and Completion}
\label{app:recursive-completion}

We now prove that, as $N\to\infty$, the probability that all fitted centers remain within unit sup-norm distance of $g_0$ throughout the recursion tends to one:
\[
 \Pr\!\left(
   \max_{0\le j\le J}\|\widehat g_j-g_0\|_\infty\le1
 \right)\longrightarrow1.
\]
Fix a sufficiently small $\delta>0$ and consider the barriers
\begin{equation}\label{eq:app-barriers}
 \max_{\ell\le j}\|\widehat g_\ell-g_0\|_\infty\le1,
 \qquad C_j\le\delta N_j^{3/5}.
\end{equation}
Section~\ref{app:startup} gives
$\|\widehat g_0-g_0\|_\infty=o_p(1)$ and
$C_0/m^{3/5}=O_p(m^{-2/5})$.
Intersect this startup event with the uniform event of Section~\ref{app:approx-stochastic}. At a first violation of either barrier, all previous centers remain localized, so based on \eqref{eq:app-sup-recursion}:
\[
 \|\widehat g_j-g_0\|_\infty
 \le C\{t^{-3/10}L(t)+\delta t^{-1/5}\}\le\tfrac12
\]
uniformly for $t\ge m$ when $m$ is large. Moreover, the pathwise definition of $C_j$ yields
\begin{equation}\label{eq:app-mass-recursion}
 \begin{aligned}
 C_j&\le C_{j-1}+n_j\|\widehat g_j-g_0\|_\infty^2\\
 &\le C_{j-1}+Cn_j\{t^{-3/5}L(t)^2+t^{-8/5}C_{j-1}^2\}\\
 &\le\delta s^{3/5}
 +n_jt^{-2/5}\{Ct^{-1/5}L(t)^2+C\delta^2\}.
 \end{aligned}
\end{equation}
Choose $\delta$ so that $C\delta^2\le\delta/5$, then with $m$ large enough and $Ct^{-1/5}L(t)^2\le\delta/5$ for every $t\ge m$. Since
$t^{3/5}-s^{3/5}\ge(3/5)n_jt^{-2/5}$,
\eqref{eq:app-mass-recursion} gives $C_j<\delta t^{3/5}$.
Thus neither barrier can first fail, and both hold along the entire path with probability tending to one.

Within these barriers,
$t^{-8/5}C_{j-1}^2\le\delta t^{-1}C_{j-1}$.
Choose $\delta$ also small enough that $\kappa=C\delta<1/5$, and then
\[
 C_j\le\left(1+\kappa\frac{n_j}{N_j}\right)C_{j-1}
 +Cn_jN_j^{-3/5}L(N_j)^2.
\]
Using
$1+x\le e^x$ and
$n_\ell/N_\ell\le\log(N_\ell/N_{\ell-1})$, we can bound the multiplicative factors from $N_a$ to $N_j$ by $(N_j/N_a)^\kappa$ based on the discrete Gronwall inequality. Integral comparison therefore gives
\begin{align}\label{eq:app-Gronwall}
 C_j
 &\le C_0(N_j/m)^\kappa
 +CN_j^\kappa\sum_{\ell=1}^j n_\ell
 N_\ell^{-3/5-\kappa}L(N_\ell)^2\nonumber\\
 &\le C_0(N_j/m)^\kappa
 +CN_j^\kappa L(N_j)^2\int_m^{N_j}x^{-3/5-\kappa}\,dx\nonumber\\
 &\le C_0(N_j/m)^\kappa+CN_j^{2/5}L(N_j)^2.
\end{align}
The startup contribution is controlled without any additional relation between $m$ and $N$:
\[
 \frac{C_0(t/m)^\kappa}{t^{1/5}}
 =\frac{C_0}{m^{1/5}}(m/t)^{1/5-\kappa}
 \le\frac{C_0}{m^{1/5}}=O_p(1),\qquad t\ge m.
\]
Consequently,
\begin{equation}\label{eq:app-C-final}
 C_J=O_p\{N^{2/5}\log^8(e+N)\}.
\end{equation}

Finally, at the deterministic endpoint $N$, the fixed-endpoint second-moment bound of Section~\ref{app:approx-stochastic} gives
$\|Z_N\|_{\alpha_N,*}=O_p(N^{-2/5})$.
Since $C_{J-1}\le C_J$, based on \eqref{eq:app-C-final}, we have
\[
 N^{-9/10}C_{J-1}
 =O_p\{N^{-1/2}\log^8(e+N)\}
 =o_p(N^{-2/5}).
\]
Also $N^{-9/10}\{1+\log(N/m)\}=o(N^{-2/5})$. Returning to \eqref{eq:app-gradient-bound}, based on strong convexity and stationarity, we deduce
$\|\widehat g_J-s_N\|_{\alpha_N}=O_p(N^{-2/5})$.
With the approximation error established in the beginning, we can deduce:
\[
 \boxed{\|\widehat g_N^{\mathrm{ren}}-g_0\|_{\alpha_N}
 =O_p(N^{-2/5}).}
\]
The $L^2$ conclusion follows from
$\|u\|_2\le\|u\|_{\alpha_N}$, completing the proof of Theorem~\ref{thm:consistency}.
\end{document}